%% file: main.tex
\documentclass{article}

\pdfoutput=1

\usepackage[preprint,nonatbib]{neurips_2024}

\usepackage[utf8]{inputenc} %
\usepackage[T1]{fontenc}    %
\usepackage{hyperref}       %
\usepackage{url}            %
\usepackage{booktabs}       %
\usepackage{amsfonts}       %
\usepackage{nicefrac}       %
\usepackage{microtype}      %
\usepackage{bbding}
\usepackage{times}
\usepackage{epsfig}
\usepackage{graphicx}
\usepackage{tabularx}
\usepackage[table,dvipsnames]{xcolor}
\usepackage{amsmath}
\usepackage{amssymb}
\usepackage{arydshln}
\usepackage[font=small]{caption}
\usepackage{bm}
\usepackage[nameinlink]{cleveref}
\usepackage{wrapfig}
\usepackage{multirow}
\usepackage{arydshln}
\usepackage{multirow}

\newcommand{\mypar}[1]{\vspace{-1mm}\paragraph{#1}}

\def\etal{\emph{et al}. }
\def\eg{\emph{e.g}. }

\definecolor{cvprblue}{rgb}{0.21,0.49,0.74}

\title{NeuroBind: Towards Unified Multimodal Representations for Neural Signals}
\hypersetup{
    colorlinks=true,
    citecolor=cvprblue,
}

\makeatletter
\def\adl@drawiv#1#2#3{%
        \hskip.5\tabcolsep
        \xleaders#3{#2.5\@tempdimb #1{1}#2.5\@tempdimb}%
                #2\z@ plus1fil minus1fil\relax
        \hskip.5\tabcolsep}
\newcommand{\cdashlinelr}[1]{%
  \noalign{\vskip\aboverulesep
           \global\let\@dashdrawstore\adl@draw
           \global\let\adl@draw\adl@drawiv}
  \cdashline{#1}
  \noalign{\global\let\adl@draw\@dashdrawstore
           \vskip\belowrulesep}}
\makeatother

\author{%
\begin{tabular}{c}
  Fengyu Yang$^{1*}$ \quad Chao Feng$^{2*}$ \quad Daniel Wang$^{1}$\thanks{equal contribution} \quad Tianye Wang$^{3}$ \\
  Ziyao Zeng$^{1}$ \quad Zhiyang Xu$^{4}$ \quad Hyoungseob Park$^{1}$ \quad Pengliang Ji$^{5}$\\
  Hanbin Zhao$^{6}$\quad Yuanning Li$^{7}$\quad Alex Wong$^{1}$
\end{tabular} \\
\begin{tabular}{c}
  $^1$Yale University \quad $^2$University of Michigan \quad $^3$Peking University \quad $^4$Virginia Tech \\
  $^5$Carnegie Mellon University $^6$ Zhejiang University $^7$ ShanghaiTech University \\
  \end{tabular}
  \texttt{}\\
}

\begin{document}

\maketitle
\input{sec/0_abstract}    
\input{sec/1_intro}
\input{sec/2_related_work}
\input{sec/3_method}
\input{sec/4_experiment}

\input{sec/5_conclusion}

\newpage
\bibliographystyle{plain}
\bibliography{main.bib}

\end{document}

%% file: sec/0_abstract.tex
\begin{abstract}
  Understanding neural activity and information representation is crucial for advancing knowledge of brain function and cognition. Neural activity, measured through techniques like electrophysiology and neuroimaging, reflects various aspects of information processing. Recent advances in deep neural networks offer new approaches to analyzing these signals using pre-trained models. However, challenges arise due to discrepancies between different neural signal modalities and the limited scale of high-quality neural data. To address these challenges, we present NeuroBind, a general representation that unifies multiple brain signal types, including EEG, fMRI, calcium imaging, and spiking data. To achieve this, we align neural signals in these image-paired neural datasets to pre-trained vision-language embeddings. Neurobind is the first model that studies different neural modalities interconnectedly and is able to leverage high-resource modality models for various neuroscience tasks. We also showed that by combining information from different neural signal modalities, NeuroBind enhances downstream performance, demonstrating the effectiveness of the complementary strengths of different neural modalities. As a result, we can leverage multiple types of neural signals mapped to the same space to improve downstream tasks, and demonstrate the complementary strengths of different neural modalities. This approach holds significant potential for advancing neuroscience research, improving AI systems, and developing neuroprosthetics and brain-computer interfaces. 
\end{abstract}

%% file: sec/1_intro.tex
\section{Introduction}
Understanding neural activity and information representation within biological neural systems is crucial for advancing knowledge of brain function and cognitive processes. Neural activity, characterized by electrical impulses, chemical signals, and oscillatory patterns, underpins behavior, perception, and cognition. Different neural signal modalities captured through recording techniques, such as electrophysiology and neuroimaging, reflect various aspects of information representation at different spatial and temporal scales \cite{uludaug2014general}. By studying these dynamic signals, researchers can decipher how information is encoded, processed, and transmitted across neural circuits. This research enhances our comprehension of neural mechanisms and has significant implications for treating neurological disorders, improving artificial intelligence, and advancing neuroprosthetics and brain-computer interfaces. Recent advances in deep neural networks, particularly for visual and auditory recognition, offer new approaches for analyzing information representation in neural signals using pre-trained deep neural networks \cite{yamins2013hierarchical,yamins2016using}. These models, used in neural encoding and decoding, predict neural activity from external stimuli \cite{millet2022toward,li2023dissecting} and decode sensory stimuli from neural activity \cite{tang2023semantic}, effectively connecting and aligning neural features with external sensory stimuli.

One major obstacle in developing pretrained deep neural network models for neural encoding and decoding analysis is the discrepancy between different signal modalities \cite{Tajbakhsh2016Convolutional}. While images, videos, speech, and text can be used to train models on vast amounts of internet-scale data, brain signals are considered an "expensive" modality because they can only be acquired from real-world subjects using complex devices. Consequently, the scale of high-quality paired neural data is relatively small. Additionally, different neural modalities cover distinct aspects of brain activity. For example, microelectrode arrays can record high-resolution spiking activity but only in a very small brain area with a limited number of neurons \cite{Du2009High-resolution,Ballini2014A}. Calcium imaging can capture whole-brain dynamics at the cellular scale, but it is challenging to apply to larger animals like primates \cite{Bollimunta2020Head-mounted, Sadakane2015Long-Term}. Electroencephalography (EEG) and functional magnetic resonance imaging (fMRI) are relatively easier to acquire but have a lower signal-to-noise ratio and limited spatiotemporal resolution.

In practice, brain signals recorded from different types of devices are commonly treated as different modalities, thus being studied independently in the current neuroscience literature\cite{mai2024brainconditional}.
To scale up the model for brain signals, we argue that these brain signals, although recorded using different methods, should be studied interconnectedly as they reflect different aspects of the same neural activity \cite{Whittingstall2013Physiological,Panzeri2008On}, where one embedding should be able to represent different modalities of signals. 
For example, in a typical controlled experiment setting, when the only varying stimuli is the visual input, then we can assume that the corresponding change in the state of neurons are caused by the change in image stimuli -- meaning the variance of the neural signals can be mostly explained away by the variance in the visual stimuli \cite{festa2021neuronal,kohn2005stimulus,Martin2013Functional}. 
This means that an effectively learned representation based on visual data should also be a sufficient representation for neural signals that correspond to visual stimuli, in arbitrary recording modality. Therefore large scale pretrained embedding spaces (\eg CLIP~\cite{radford2021learning}) have the potential to be used as a feature space shared by different neural modalities.
 
In this work, we present \textbf{NeuroBind}, which learns a general representation based on pretrained image embedding space that unifies multiple types of brain signals including EEG, fMRI, calcium imaging, and spiking data. We leverage the above typical controlled experimental setting by aligning different types of brain signals with pretrained vision-language embeddings~\cite{lee2022sound,Girdhar2023ImageBindOE, Xue2022ULIPLA} from image-paired brain datasets. 
Thus, we are able to take advantage of the complementary strength across different brain modalities for better downstream applications. 
In addition, by binding various neural signal representations with high-resource modalities like vision and language, we are able to leverage powerful off-the-shelf models trained on these modalities (\eg LLM, text-based diffusion) in neuroscience research. To the best of our knowledge, this is the first model for neural data that works across multiple modalities of neural recording signals.

Our model is capable of conducting a variety of tasks related to various neural signal modalities, some of which have been rarely analyzed via deep learning (\eg calcium imaging), as illustrated in Figure 1: 
(i) We are the first to conduct cross-modal retrieval between different neural signals. (ii) We apply our model to the zero-shot neural signal classification task and find that we can achieve better performance by combining features from other neural modalities via retrieval. (iii) We apply it to zero-shot image reconstruction tasks with off-the-shelf text-to-image diffusion models, where we are the first to reconstruct images from multiple neural signals. (iv) We combine our model with large language models (LLM), allowing us to achieve a more comprehensive understanding of various neural signals via language.

%% file: sec/2_related_work.tex
\section{Related Work}

\mypar{DNN-based visual neural encoding models.} 
Understanding how the brain processes and represents information from the sensory system has always been a central task in neural science. With the recent advancements of neural networks (NN), researchers have explored its power in the realm of neuroscience, many utilizing the ANN for predicting how neurons or neural populations encode sensory stimuli \cite{yamins2016using}. Yamins \etal \cite{yamins2013hierarchical} were one of the first researchers to introduce CNN to the task of predicting visual cortex neuronal responses from image stimuli. Kell \etal \cite{Kell2018ATN} compared the features of a learned neural network and mapped it to recorded brain signals, showing that a task-optimized NN model can be similar to the human cortical organization. Khosla \etal \cite{khosla2021cortical} showed that cortical response to naturalistic videos can be predicted using features extracted from convolution and recurrent networks. Zheng \etal \cite{zheng2021unraveling} showed that retinal neuronal activity can be modeled and predicted by convolution and recurrent neural networks. 

\mypar{Visual neural signal decoding.}
Apart from predicting the neural activity through learning the encoding process of neurons, another important direction is to learn models that decode information about the stimuli given the neural responses \cite{mai2024brainconditional}. Many earlier works have focused on decoding simple patterns \cite{kay2008identifying}, where simpler techniques such as linear mapping can be sufficient for both classification and or reconstruction. For natural images that contain complex features and shapes, decoding is usually considered a harder task. Although results have shown that a simple probabilistic model can decode and reconstruct natural images from neural activity \cite{nishimoto2011reconstructing, Yoshida2020NaturalIA}, researchers have opted to use artificial neural networks for natural image decoding. The pioneering work of \cite{Horikawa2015GenericDO} introduced CNN networks as a feature extractor, and trained linear mapping models from neural signals to pre-trained CNN features, which allows them to retrieve the class of the image stimuli by matching the category-average features with the predicted features. \cite{Wen2016NeuralEA} also used linear mapping to decode fMRI signals to estimate low-level visual features extracted by the first CNN layer, and used a deconvolutional network to reconstruct the image stimuli from those features. 

\mypar{Visual decoding with generative methods.}
Inspired by the success of generative models in computer vision, researchers have explored DNN-based generative models as neural decoders. These models leverage large-scale image data, providing a strong prior on naturalistic images. GAN models have been used for direct reconstruction from neural signals~\cite{seeliger2018generative, Huang2020PerceptiontoImageRN, Mishra2022NeuroGANIR,Kavasidis2017Brain2ImageCB}. Seeliger \etal\cite{seeliger2018generative} trained a GAN on image stimuli paired with neural recordings, linearly mapping neural signals to the GAN's latent space for image reconstruction. Huang \etal \cite{Huang2020PerceptiontoImageRN} directly trained a GAN model conditioned on neural signal features. Lin \etal \cite{lin2022mind} used a StyleGAN conditioned on fMRI signals mapped to CLIP space. Mishra \etal \cite{Mishra2022NeuroGANIR} proposed an attention-based NeuroGAN, extending GAN usage to EEG signals by conditioning the network on neural signals and employing a pre-trained classifier to generate class-specific images. Recently, the success of diffusion models~\cite{yang2023generating,dou2024tactile,rombach2022high,ho2020denoising,song2020denoising,ho2022video,blattmann2023stable,poole2022dreamfusion} in computer vision has led researchers to explore their use for image reconstruction from neural recordings, significantly improving image quality ~\cite{Takagi2023HighresolutionIR, ozcelik2023natural, Chen2023CinematicMH, Lu2023MindDiffuserCI, Scotti2023ReconstructingTM,liu2023brainclip}. Takagi \etal \cite{takagi2023high} used a latent diffusion model to reconstruct perceived images from fMRI signals recorded in visual and semantic areas. Ozcel \etal \cite{ozcelik2023natural} employed a two-stage generation pipeline, combining coarse reconstruction from a VDVAE with CLIP-aligned neural signal features to condition a diffusion model for fine-detailed reconstructions. Chen \etal \cite{Chen2023CinematicMH} aligned neural features to CLIP embeddings via contrastive pre-training and applied the diffusion model to fMRI data frames for video reconstruction. While these approaches generate high-resolution images with correct semantic classes, they often lack structural detail compared to the ground truth images. To address this, Lu \etal \cite{Lu2023MindDiffuserCI} added an image structural refinement step aligning generated images to the ground truth's structural layout. Scotti \etal \cite{Scotti2023ReconstructingTM} used a two-stage training process where fMRI signals were first matched to stable diffusion encoder embeddings for low-level reconstruction, then upscaled with a separate diffusion model conditioned on neural signals matched with CLIP features. Aligning neural data to pre-trained embedding spaces like CLIP \cite{radford2021learning} is common in reconstruction pipelines. However, current research has not explored concurrently aligning different neural modalities onto the same feature space. Our model addresses this gap, enabling reconstruction from arbitrarily aligned neural signal sources without needing to train the generative model for each modality specifically.
\input{figure/method}

\mypar{Representation learning for neural signals.} 
Recent advances in data-driven machine learning and the increasing availability of large-scale data have encouraged researchers to explore neural feature representations better linked to sensory stimuli and behavior \cite{urai2022large}. Classic techniques such as principal component analysis (PCA), t-SNE, and UMAP can effectively reduce the dimensionality of neural data while retaining information \cite{van2008visualizing, mcinnes2018umap}. However, these simple unsupervised methods cannot effectively utilize information from external data \cite{Schneider_2023}. Deep learning-based methods for representation learning \cite{han2023imagebind, radford2021learning} have driven lots of applications across various data modalities \cite{zhu2023pointclip, zhang2022can, chen2023iquery, zhang2021dspoint, zeng2024wordepth, lao2022depth, dou2024tactile, yang2022touch, yang2023generating, Wu_2023_boosting, Zhao2022RBCRT,yang2022sparse, li-19-segmentation, li-23-deception-detection, li-24-vqa,xu2024vision,lin2024optimalsamplinglearningsdf,lin2023patch}, and their applications in neural data have also been deeply explored, such as goal-driven classification networks \cite{Wonjun_EEG_representation} and variational autoencoders \cite{sani2021modeling,zhou2020learning}. Recently, contrastive learning frameworks, inspired by SimCLR \cite{chen2020simple}, have been adapted for neural data \cite{Schneider_2023,Lei2023ViTLens2GT,han2023onellm,mai2024brainconditional}. CEBRA \cite{Schneider_2023} applied contrastive learning using behavior labels and time as supervision, while VIT-Lens2 \cite{Lei2023ViTLens2GT} and one-LLM \cite{han2023onellm} used contrastive learning on fMRI and EEG signals paired with images to align with other modalities, such as audio and 3d point clouds. CLIP-MUSED \cite{Zhou2024CLIPMUSEDCM} learned a joint embedding space across different subjects for decoding via classification. Our work utilizes a similar contrastive learning objective \cite{Schneider_2023,Lei2023ViTLens2GT,han2023imagebind,yang2024binding, qiu2021vt,yang2022touch,feng2023self,yang2022sparse} to align cross-modality neural recordings to a single feature space based on the CLIP image encoder. We demonstrate that without fine-tuning the embedding space (keeping the image encoder frozen), our model performs well on many downstream tasks with zero-shot ability, proving that using vision stimuli as external supervision can effectively learn neural representations, comparable to behavioral or time labels used in CEBRA \cite{Schneider_2023}.

%% file: figure/method.tex
\begin{figure*}[!t]
    \centering
    \includegraphics[width=\linewidth]{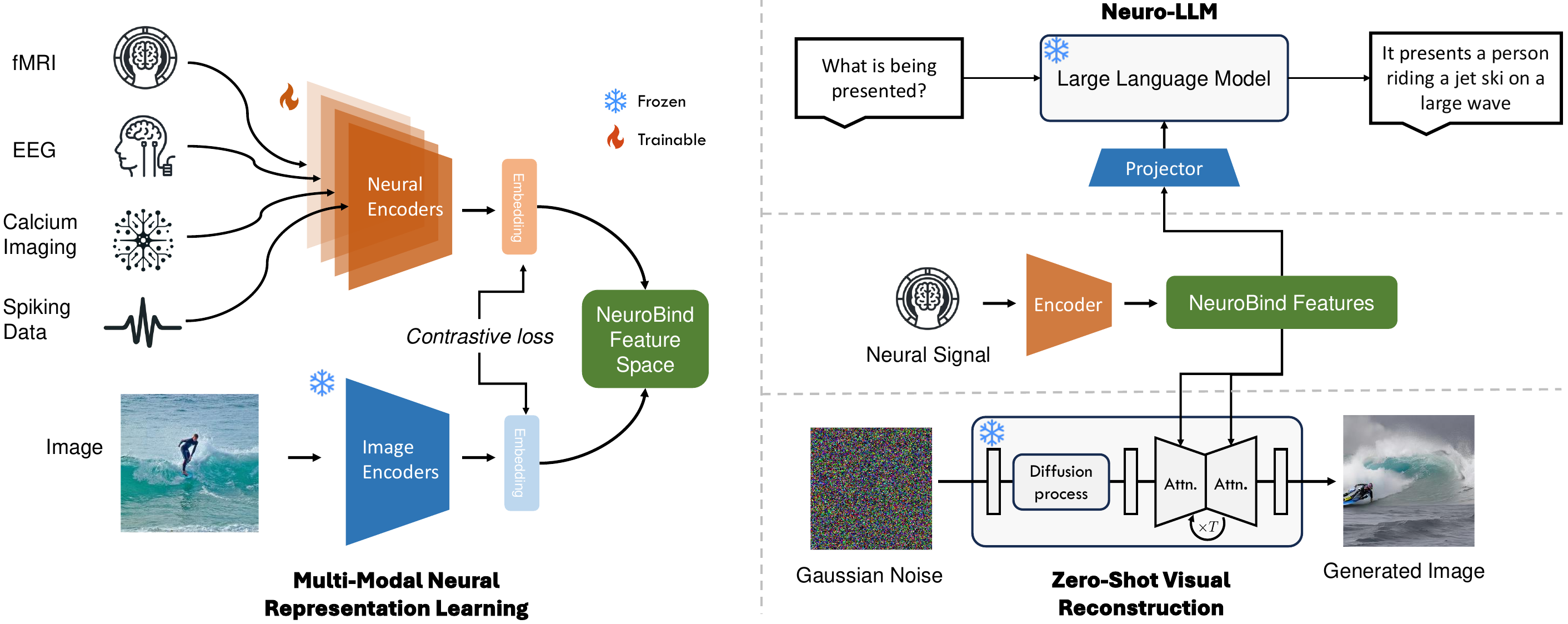}
    \caption{{\bf Method Overview.} (left) We align embeddings from neural signals with a pre-trained visual embedding trained on large-scale vision language datasets. (right) Our Neurobind embeddings can be applied in pre-trained LLM and text-based diffusion models for neuroscience tasks without re-training.} 
    \label{fig:method}
\end{figure*}

%% file: sec/3_method.tex
\newcommand{\bv}{{\mathbf v}}
\newcommand{\bn}{{\mathbf n}}
\newcommand{\fn}{\mathcal{F}_N}
\newcommand{\fv}{\mathcal{F}_V}

\section{Method}
Our goal is to develop unified neural representations for various brain signal modalities, including EEG, fMRI, calcium imaging, and spiking data. We achieve it by mapping these brain signals into latent space where their embedding is aligned with those of the corresponding visual stimuli from pre-trained vision-language models~\cite{han2023imagebind, radford2021learning}. In this section, we first introduce our visual-neural pre-training inspired by~\cite{han2023imagebind, radford2021learning}, which enables the emergence of interconnections between brain signals and other modalities like language (Figure~\ref{fig:method} (left)). We then introduce several downstream application tasks empowered by our proposed pre-training alignment model.

\subsection{Binding brain signals with vision and language}
We focus on learning a neural representation that captures correspondences between different brain signals and external modalities (e.g. vision, text), which is useful for downstream tasks. While most of the datasets only involve brain signals and their corresponding visual stimuli, we align the brain embedding to a pre-trained image embedding, which is in a latent space connected with other modalities trained from large-scale image-paired datasets~\cite{Girdhar2023ImageBindOE}. 

Specifically, we denote ($\mathcal{V}$, $\mathcal{N}$) as the pair of modalities where $\mathcal{N}$ as the neural modality and $\mathcal{V}$ as the corresponding visual stimuli. 
Thus, given a mini-batch containing $B$ paired neural and visual signal $\{(\bn_i, \bv_i)\}_{i=1}^B$, we first encode them into L2-normalized embeddings $\fn(\bn_i) \in \mathbb{R}^{D}$ and $\fv(\bv_i)\in \mathbb{R}^{D}$, where $\fn$ is the trainable neural encoder for the corresponding neural modality and $\fv$ is a frozen pre-trained visual encoder. Then, we maximize the probability of finding the corresponding neural-visual pair in the batch of $B$ samples using InfoNCE loss~\cite{oord2018representation}. To maximize the agreement between a neural signal and its corresponding visual stimuli, the loss can be written as the following:
\begin{equation}\label{eq1}
  \mathcal{L}_{N, V} = -\frac{1}{B}\sum_{i = 1}^{B}{\log}\frac{\exp(\fv(\bv_i) \cdot \fn(\bn_i)/\tau)}
  {\sum_{j = 1}^{B} {\exp}(\fv(\bv_i) \cdot \fn(\bn_j)/\tau)} \text{,}
\end{equation}

where $\tau$ is a temperature hyperparameter~\cite{wu2018unsupervised} that controls the smoothness of the softmax distribution. Analogously, we define the loss $\mathcal{L}_{V, N}$ that matches the visual stimuli to the neural example. Thus we minimize the overall loss function:
\begin{equation}\label{eq2}
    \mathcal{L} = \mathcal{L}_{V, N} + \mathcal{L}_{N, V}
\end{equation}

Applying InfoNCE loss~\cite{oord2018representation} will pull the corresponding neural-visual pair close together and push it away from the other pairs. Since the pre-trained visual embeddings have already been aligned with the text embeddings, we thus obtain a multimodal representation of brain signals.

\subsection{Applications}
After aligning neural embeddings with pretrained visual embedding, we implicitly build connections between neural signals to other modalities like language. Since visual and language embeddings are pre-trained from large-scale datasets~\cite{Girdhar2023ImageBindOE}, we are able to leverage their emergent ability to conduct zero-shot and cross-modal applications below.

\mypar{Cross-modal retrieval across neural modalities.}
Leveraging vision as an intermediate modality, we are able to build connections among different neural signals. Thus, the goal of the cross-modal retrieval task is to find the two neural signals from different modalities that share the same or similar visual stimuli. This is done by comparing the cosine similarities between the neural embedding from one modality and those of the candidate embeddings from another neural modality.

\mypar{Zero-shot brain signal classification.}
Vision-language pre-trained models like CLIP~\cite{radford2021learning} have demonstrated a strong emergent ability to conduct zero-shot classifications via prompting. Since our neural representations are implicitly aligned with CLIP language embeddings, NeuroBind is capable of conducting brain signal classification for the corresponding visual stimuli. Specifically, we first encode brain signals and manually designed textual prompts containing the names of different categories. Then, we compute the cosine similarity between neural and textual embeddings and select the category with the highest score.

\mypar{Zero-shot image reconstruction from neural signals.}
Reconstructing visual stimuli from measured neural signals has been a meaningful and challenging task. However, prior works mostly focus on neural signals that are easy to acquire, like fMRI or EEG. This task is mostly done by training costly generative models conditional on these neural signals. As we have already aligned neural signals to text, we open up the possibility of leveraging the pretrained text-to-image diffusion~\cite{ramesh2022hierarchical} by replacing the condition of the text embedding with our neural embeddings to conduct zero-shot image reconstruction from various neural signals (Figure~\ref{fig:method} (right top)).

\mypar{Neuro-LLM.}
Binding neural signals with language in the latent space actually allows us to interpret and describe neural signals explicitly. We are able to create our own Neuro-LLM via an existing vision-language LLM~\cite{zhang2023llama,gao2023llama} with the vision embedding aligned with neural embedding (Figure~\ref{fig:method} (right bottom). By switching the image encoder to the neural encoder from various neural signals, we can describe the visual stimuli paired with the corresponding neural signals via question-answering.

%% file: sec/4_experiment.tex
\section{Experiments}

\subsection{Implementation Details}

We base the frozen image encoder of our neurobind model on~\cite{Girdhar2023ImageBindOE}. We adopt Vision Transformer (ViT)~\cite{dosovitskiy2021an} as the backbone for all neural encoders, which contains 24 multi-head attention blocks with 8 heads on each with an output dimension of 512. We use the AdamW optimizer~\cite{kingma2015adam,loshchilov2017decoupled} as the optimizers for all modalities. All the models are trained on 4 Nvidia A6000 GPUS. For fMRI signals and EEG signals, we train our model with the batch size of 64 on each GPU for 200 epochs. We set the base learning rate of $2 \times 10^{-4}$ and the cosine decay learning rate scheduler of $0.2$. For Calcium Imaging signals, we train the encoder with the same batch size and learning rate as above but with 100 epochs. For microelectrode arrays, we set the batch size to be 32 and base learning rate as $5 \times 10^{-4}$ for 400 epochs. 

\subsection{Datasets}

\mypar{Natural Scenes Dataset.} For the fMRI data, we used the Natural Scenes Dataset (NSD)~\cite{allen2022massive}, which is a comprehensive high-resolution (1.8-mm) whole-brain 7T fMRI dataset. It consists of data collected from eight human participants, each viewing 9,000–10,000 unique color natural scenes during 30–40 scan sessions. The total dataset includes responses to 70,566 distinct images from COCO \cite{lin2015microsoft}, which contains complex natural scenes with rich semantic information. Specifically, we used the data from subject 1, which includes 8859 image stimuli for the training set and 982 image stimuli for the test set. We averaged the fMRI responses that had multiple trials over one image. We also utilized the captions that correspond to the images from COCO, which is used for our CLIP score evaluation.

\mypar{EEG ImageNet Dataset.} Regarding EEG data, we used the EEG ImageNet dataset from \cite{palazzo2020decoding}. The dataset includes EEG recordings from six human subjects, each viewing 2,000 images from 40 object categories, sourced from the semantic-rich ImageNet dataset, with a (1600,200,200) split for training, validation, and testing. Each category contains 50 distinct images, resulting in a total of 12,000 EEG sequences. We specifically used data from subject 1. These recordings were captured at a 1 kHz sampling rate across 128 channels using a Brainvision EEG system. Class labels from ImageNet corresponding to these images were also for classification tasks.

\mypar{V4 Widefield Calcium Imaging Dataset.} For Calcium imaging data, we used the V4 data from \cite{wang2023large}. This dataset comprises widefield calcium imaging recordings from the dorsal V4 region of three awake macaque monkeys, captured in response to over 20,000 color natural images. The calcium indicator GCaMP5G was expressed in V4 cortical area in the visual cortex, and a 10mm-diameter optical window was used for imaging.The dataset includes both a training set of cortical responses to 17,000-20,000 natural images. Our training and experiments used data from monkey subject 1, with a total of 20,000 training images. Additionally, there is a validation set with responses to an additional 500 natural images, where the stimuli were repeated ten times in a randomly interleaved manner to ensure robustness. The responses to the 10 trials were then averaged during inference.

\mypar{V1 Spiking Dataset.} As for spiking data, we used the primary visual cortex (V1) data from \cite{Cadena2017DeepCM}, which comprises electrophysiological recordings from the V1 of two awake, fixating rhesus macaques. The neural activity was captured using a 32-channel linear silicon probe across 17 recording sessions, resulting in data from 166 isolated neurons. The stimuli set included 1,450 grayscale natural images sourced and cropped from ImageNet, alongside four texturized versions of each image generated using a parametric model for texture synthesis. This resulted in a total of 7,250 images, each displayed for 60 ms without blanks in between. The images were masked to cover 2 degrees of the visual field and randomized to ensure diverse exposure. We also used a coloring model from \cite{zhang2016colorful,zhang2017real} to retrieve the colorized stimuli from the dataset.
\subsection{Results}
\input{tables/retrieval}
\mypar{Cross-modal retrieval with neural signals.} We test how well different neural representations are aligned in our model via the cross-modal retrieval task. Given one neural signal, our goal is to identify the corresponding neural signals from other neural modalities that describe the same visual stimuli. We evaluate our method on all four neural datasets. However, as the visual stimuli from these datasets are not fully identical, we measure the similarity of the corresponding visual stimuli of the retrieved neural signals. The similarity between two visual stimuli is measured via the cosine distance of their CLIP score. We compare our method with four supervised cross-modal retrieval baselines including Canonical Correlation Analysis (CCA)~\cite{Hotelling1936RelationsBT}, Partial Least Squares (PLSCA)~\cite{Jong2001CanonicalPL}, Deep Aligned Representations (DAR)~\cite{Aytar2017SeeHA}, and Deep Supervised Cross-Modal Retrieval (DSCMR)~\cite{Zhen2019DeepSC}. Since our model is directly not trained on paired neural modalities, we see our model as unsupervised in this task.
The results are shown in Table~\ref{tab:retrieval_results}. NeuroBind significantly outperforms existing supervised methods, which demonstrates strong emergent alignment across different neural modalities. 

\input{figure/fmri_qualitative}
\mypar{Zero-shot Brain Signal Classification.} Given a brain signal, our goal is to infer the semantic category of its corresponding visual stimuli without retraining, which is measured via accuracy. 
\input{tables/cls}
We conduct zero-shot brain signal classification on EEG ImageNet~\cite{palazzo2020decoding} by prompting the model with ``This is a brain signal of [CLS]'', where [CLS] is the name of the class label. We compare our method with supervised linear probing from scratch and the pertaining method from~\cite{bai2023dreamdiffusion}, where we use its released EEG encoder and train a linear layer upon it. We show our zero-shot classification performance in Table~\ref{tab:cls}. We can see clearly our zero-shot performance is even slightly better than the supervised linear probing method. We also report linear probing performance using our pre-trained encoder. Our supervised performance significantly surpassed other pretrained methods, which further demonstrates the effectiveness of our neural embeddings. 

\mypar{Retrieval Augmentation for Neural Signals.}
\input{tables/ablation}
As different types of brain signals cover distinct aspects of the brain activity, we evaluate whether their complementary information is able to help downstream applications. We test our hypothesis on the brain signal classification task for EEG signal. We select our baseline as the linear probing performance on EEG ImageNet dataset. Then, we retrieve the closest embedding from other neural modalities and train a linear layer on top of the concatenated features. We show our results in Table~\ref{tab:ablation}. The performance is improved by 1.6$\%$ when retrived fMRI embedding and 1.6$\%$, 1.6$\%$ for calcium imaging and spiking data embeddings. It achieves an overall $3.8\%$ when all four embeddings are added.

\mypar{Image Reconstructions from Neural Signals.}

\input{tables/generation}
In this part, we demonstrate that we can combine our neural embeddings with an off-the-shelf image synthesis model easily to perform the image synthesis tasks conditioning various neural signals in a zero-shot manner. Concretely, we use pretrained frozen text-to-image stable diffusion models~\cite{ramesh2022hierarchical} conditioned on fMRI, EEG, and calcium imaging (CI) to perform image reconstruction. As shown in Table~\ref{tab:generation}, our method can outperform baselines by a relatively large margin in terms of FID~\cite{heusel2017gans}, CLIP Score~\cite{ramesh2022hierarchical}, and Inception Score~\cite{salimans2016improved}. Moreover, we present some qualitative results in Figure~\ref{fig:fMRI},~\ref{fig:eeg},~\ref{fig:CI}. It indicates that our method could generate semantically consistent images in a zero-shot manner. 
\input{figure/eeg_qualitative}
\input{figure/ci_qualitative}
\input{figure/neuro_llm}
\mypar{Neuro-LLM.} 
Interpreting neural signals, crucial for understanding brain functions, is naturally challenging due to human perceptual limitations. 
To address this, we integrate our neuro embeddings into a large language model (LLM), leveraging its robust understanding and reasoning capabilities for neural signals interpretation, and name it Neuro-LLM. 
\input{tables/llm}
Neuro-LLM is capable of describing and understanding the scene of visual stimuli, which is non-trivial to humans, demonstrating the usefulness of combining neural signals with LLMs. 
We show some example tasks in Figure~\ref{fig:neuro-llm}. Quantitatively, we compare our model with two open-source vision-language models (VLMs): ImageBind-LLM~\cite{han2023imagebind} and One-LLM~\cite{han2023onellm} in the fMRI signal captioning task by feeding them the same fMRI signals and text prompts. We evaluate all methods on the test split of NSD~\cite{allen2022massive}. Following prior work~\cite{han2023onellm}, we use CIDEr~\cite{vedantam2015cider} and ROUGE-L~\cite{lin2004rouge} as evaluation metrics. As shown in Table~\ref{tab:caption}, our Neuro-LLM outperforms other VLMs by a large margin, indicating that our Neuro-LLM has much better understanding capabilities for fMRI signals.

%% file: tables/retrieval.tex
\begin{table}[t!]

\centering
\footnotesize
\begin{tabular}{clcccc}
\toprule
& \multirow{2}{*}{\textbf{Method}} & \multicolumn{4}{c}{\textbf{CLIP Score}}\\ 
\cmidrule{3-6}
&  &  fMRI $\leftrightarrow$ EEG & EEG $\leftrightarrow$ CI & CI $\leftrightarrow$ Spiking Data & Spiking Data $\leftrightarrow$ fMRI\\ \midrule
 \multirow{4}{*}{\shortstack[c]{Fully \\ supervised}} & CCA & 0.15 & 0.17 & 0.23 & 0.15\\
 & PLSCA & 0.23 & 0.21 & 0.25& 0.31\\
 & DSCMR & 0.37 & 0.32& 0.37 &0.28\\
 & DAR & 0.35 & 0.37 & 0.41 &0.31\\
 \cdashlinelr{1-6} 
 \multirow{1}{*}{Zero-shot} 
 & Ours& \textbf{0.50} & \textbf{0.48} & \textbf{0.55} & \textbf{0.44}\\
\bottomrule
\end{tabular}
\vspace{2mm}
\caption{
\textbf{Cross-modal retrieval across neural signals.} We evaluate the performance via computing the cosine similarity of the CLIP visual features corresponding to the retrieved neural features.} %
\label{tab:retrieval_results}
\end{table}

%% file: figure/fmri_qualitative.tex
\begin{figure*}[t!]
    \centering
    \includegraphics[width=\linewidth]{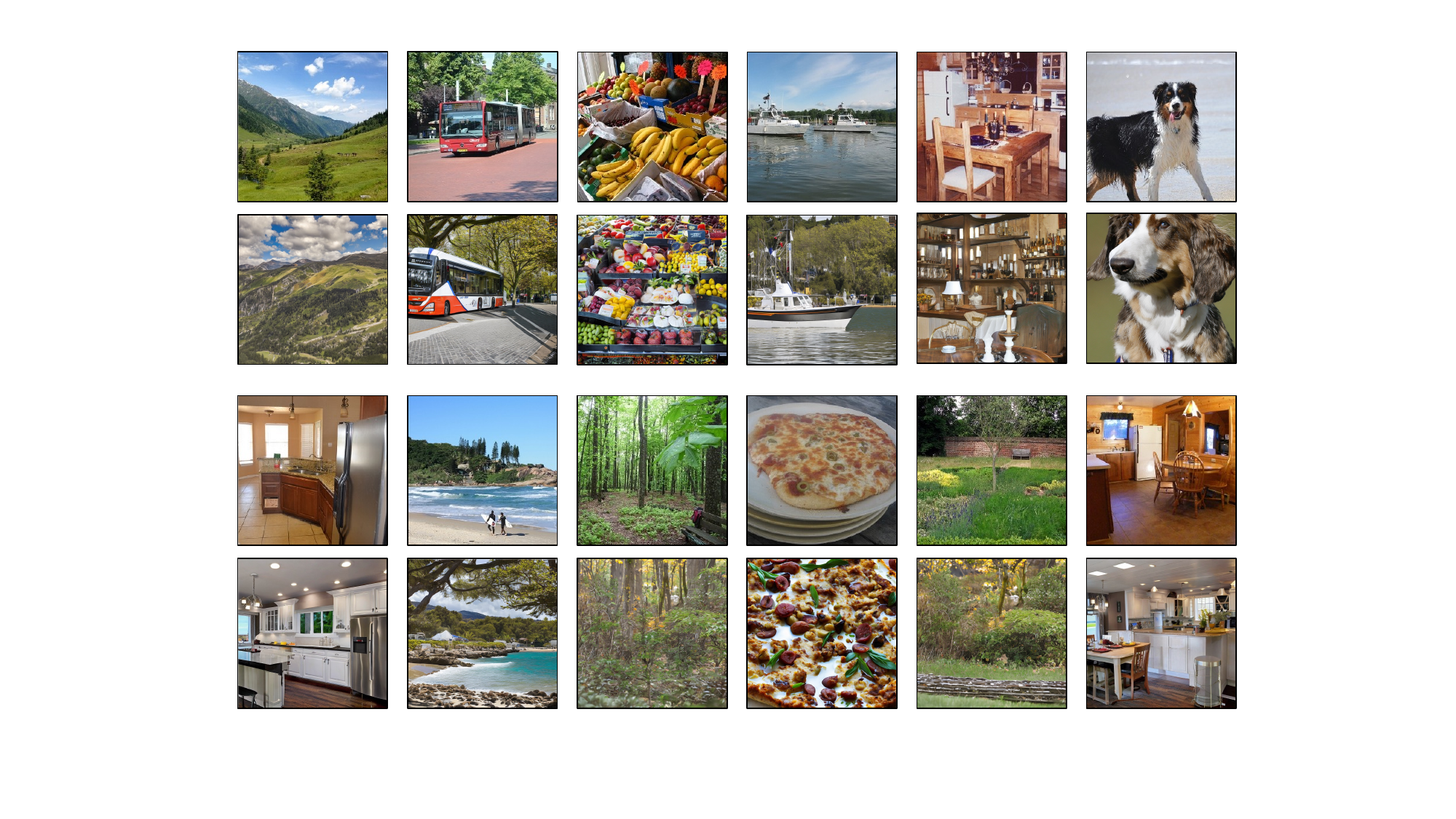}
    \caption{{\bf fMRI to image reconstruction.} We present some generated images from pretrained frozen text-to-image stable diffusion models~\cite{ramesh2022hierarchical} conditioned on our fMRI embedding. To compare, We also present the generated images that correspond to the same fMRI signal from MindDiffuser~\cite{Lu2023MindDiffuserCI}.} 
    \label{fig:fMRI}
\end{figure*}

%% file: tables/cls.tex
\begin{wraptable}{r}{7cm}
\footnotesize

\centering
\begin{tabular}{llc}
\toprule
 & \multirow{2}{*}{\textbf{Pre-training}} & \textbf{Eval} \\ \cmidrule{3-3}
&  & \emph{EEG ImageNet}\\ \midrule
Chance & -- & 2.5 \\ 
\midrule
\multirow{3}{*}{\shortstack[c]{Linear \\ Probing}} & None & 14.8 \\
& DreamDiffusion~\cite{bai2023dreamdiffusion} & 19.2 \\
& Ours & \textbf{32.7} \\
\midrule
Zero-shot & Ours  &\textbf{20.3}\\
\bottomrule
\end{tabular}
\caption{\textbf{Brain signal classification.} We compare our feature with other methods on EEG ImageNet~\cite{palazzo2020decoding} for zero-shot classification. The evaluation metric is accuracy ($\%$).} 
\vspace{-4mm}
\label{tab:cls}
\end{wraptable} 

%% file: tables/ablation.tex
\vspace{-2mm}
\begin{wraptable}{r}{6cm}
\footnotesize

\centering
\begin{tabular}{llc}
\toprule
\multirow{2}{*}{\textbf{Method}} & \multirow{2}{*}{\shortstack[c]{\textbf{Modality} \\ \textbf{Augmentation}}} & \textbf{Eval} \\ \cmidrule{3-3}
&  & \emph{EEG ImageNet}\\ \midrule
\multirow{5}{*}{\shortstack[c]{Linear \\ Probing}} & None & 32.7 \\
& fMRI & 35.3 \\
& CI & 33.9 \\
& Spiking Data & 33.5 \\
& All & \textbf{36.5} \\
\bottomrule
\end{tabular}
\caption{\textbf{Retrieval Augmentation for Brain Signal Classification.} }
\vspace{-5mm}
\label{tab:ablation}
\end{wraptable} 

%% file: tables/generation.tex
\begin{wraptable}{r}{9cm}
\centering

\footnotesize
\begin{tabular}{lccc}
\toprule
\multirow{2}{*}{\textbf{Method}}          & \multicolumn{3}{c}{\textbf{Eval}}  \\
\cmidrule{2-4}
&  FID ($\downarrow$) & CLIP ($\uparrow$) & Inception ($\uparrow$)\\
\midrule
MindDiffuser~(fMRI)~\cite{Lu2023MindDiffuserCI}       &    115.9     & 20.0 & 4.2 \\
Ours~(fMRI) &    \textbf{64.7}  & \textbf{26.1} &\textbf{5.7}\\
 \cdashlinelr{1-4} 
DreamDiffusion~(EEG)~\cite{bai2023dreamdiffusion}      &     211.7 &  20.5  &3.8\\
Ours~(EEG) &    \textbf{166.9}  & \textbf{25.3} &\textbf{4.5}\\
 \cdashlinelr{1-4} 
Ours~(CI) & 201.1 & N/A & 5.6 \\
\bottomrule
\end{tabular}
\caption{\textbf{Image reconstruction comparison.} We evaluate our visual stimuli reconstruction results for neural signals of fMRI, EEG, and CI in terms of FID~\cite{heusel2017gans}, CLIP score~\cite{ramesh2022hierarchical}, and Inception score~\cite{salimans2016improved}. It should be noted that our method is zero-shot. The FID score of CI is high because our reconstruction model wasn't fine-tuned for this specific dataset with masked peripherals.}
\vspace{-2mm} 
\label{tab:generation}
\end{wraptable}

%% file: figure/eeg_qualitative.tex
\begin{figure*}[t!]
    \centering
    \includegraphics[width=\linewidth]{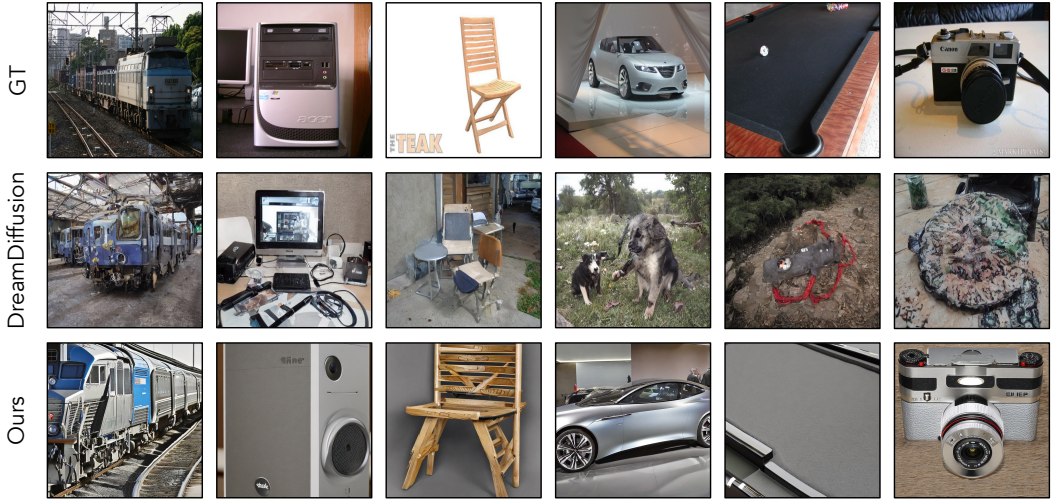}
    \caption{{\bf EEG to image reconstruction.} We show some sampled images from pretrained frozen text-to-image stable diffusion models~\cite{ramesh2022hierarchical} conditioned on our EEG embedding. For comparison, we also present the generated images corresponding to the same EEG signal from DreamDiffusion~\cite{bai2023dreamdiffusion}} 
    \label{fig:eeg}
\end{figure*}

%% file: figure/ci_qualitative.tex
\begin{figure*}[t!]
    \centering
    \includegraphics[width=\linewidth]{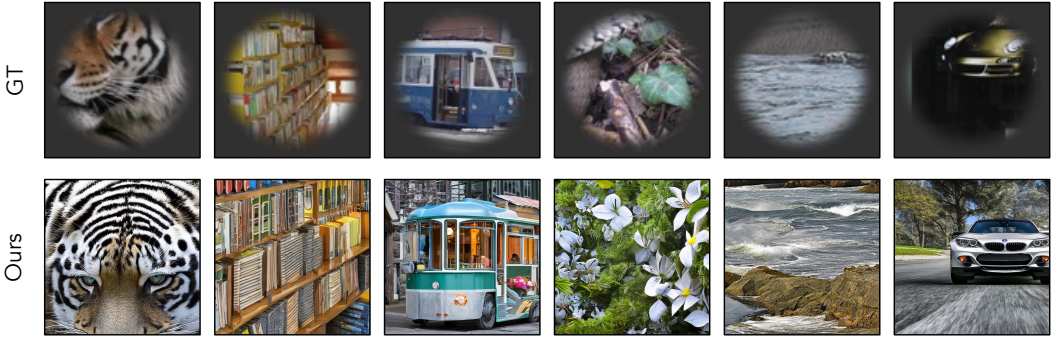}
    \caption{{\bf Calcium Imaging (CI) to image reconstruction.} We present representative images from pretrained frozen text-to-image stable diffusion ~\cite{ramesh2022hierarchical} conditioned on our CI embedding, which we are the first to conduct. } 
    \label{fig:CI}
    \vspace{-3mm}
\end{figure*}

%% file: figure/neuro_llm.tex
\begin{figure*}[t]
    \centering
    \includegraphics[width=\linewidth]{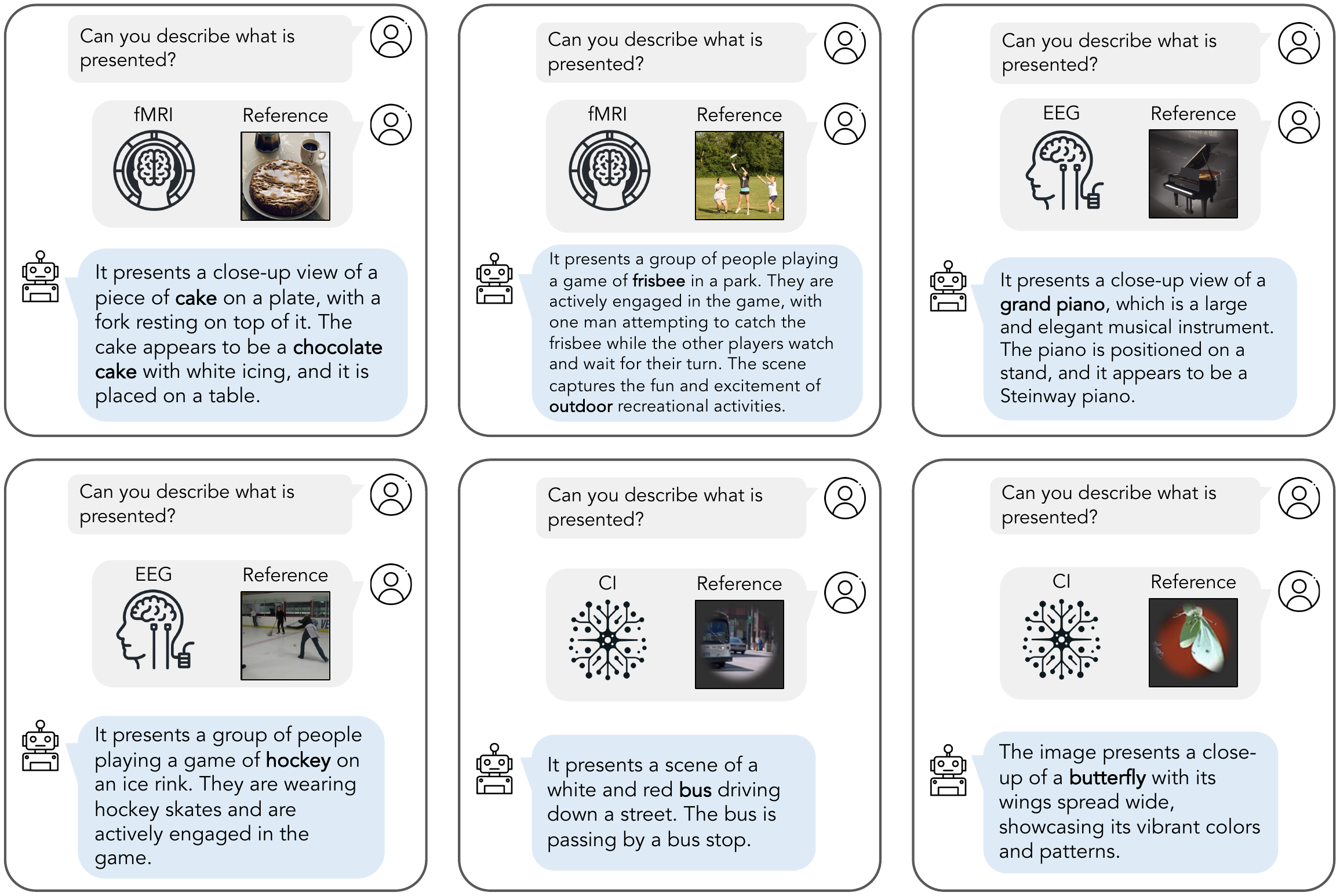}
    \caption{{\bf Neuro-LLM.} Our Neuro-LLM can understand various kinds of neural signals and describe the scene of visual stimuli. We also show reference RGB images for better demonstration.} 
    \vspace{-3mm}
    \label{fig:neuro-llm}
\end{figure*}

%% file: tables/llm.tex
\begin{wraptable}{r}{9.8cm}
\centering
\footnotesize
\begin{tabular}{llcc}
\toprule
\multirow{2}{*}{\textbf{Method}}    &   \multirow{2}{*}{\textbf{LLM}}      & \multicolumn{2}{c}{\textbf{Eval}}  \\
\cmidrule{3-4}
& & CIDEr ($\uparrow$) & ROUGE-L ($\uparrow$)\\
\midrule
ImageBind-LLM~\cite{han2023imagebind}     &  LLaMA-7B~\cite{touvron2023llama}  &    1.2     & 16.3  \\
One-LLM~\cite{han2023onellm}     &  LLaMA2-7B~\cite{Touvron2023Llama2O}  &         31.7 &  25.1   \\
 \cdashlinelr{1-4} 
Neuro-LLM (ours) &  LLaMA-7B~\cite{touvron2023llama} &  \textbf{57.8}  & \textbf{25.4} \\
\bottomrule
\end{tabular}
\caption{\textbf{Neural signals caption evaluation.} We test our Neuro-LLM and two baselines on test split of NSD~\cite{allen2022massive}.} 
\label{tab:caption}
\vspace{-4mm}
\end{wraptable}

%% file: sec/5_conclusion.tex
\vspace{-5mm}
\section{Discussion}
\vspace{-4mm}
In this work, we present NeuroBind, a multimodal neural foundation model that unifies multiple brain signal modalities, including EEG, fMRI, calcium imaging, and spiking data, and connects them with external modalities like image and text. We achieve this by aligning each modality with pre-trained vision-language embeddings of their corresponding visual stimuli. NeuroBind enables multiple brain-related tasks in zero-shot and bridges connections between different brain signal modalities that were previously studied independently, significantly advancing neuroscience research. We view our work as a substantial step toward a more comprehensive understanding of human brain activities.
\vspace{-2mm}

\textbf{Broad impact.} Our work enables cross modality study of brain function and cognitive process. Barring malicious users, we do not foresee any negative societal impact.

\vspace{-2mm}
\textbf{Limitations.} 
While our work covers multiple brain signals, there exists many other modalities that we have not explored due to limited computing resources.

%% file: main.bbl
\begin{thebibliography}{100}

\bibitem{allen2022massive}
Emily~J Allen, Ghislain St-Yves, Yihan Wu, Jesse~L Breedlove, Jacob~S Prince, Logan~T Dowdle, Matthias Nau, Brad Caron, Franco Pestilli, Ian Charest, et~al.
\newblock A massive 7t fmri dataset to bridge cognitive neuroscience and artificial intelligence.
\newblock {\em Nature neuroscience}, 25(1):116--126, 2022.

\bibitem{Aytar2017SeeHA}
Yusuf Aytar, Carl Vondrick, and Antonio Torralba.
\newblock See, hear, and read: Deep aligned representations.
\newblock {\em ArXiv}, abs/1706.00932, 2017.

\bibitem{bai2023dreamdiffusion}
Yunpeng Bai, Xintao Wang, Yan pei Cao, Yixiao Ge, Chun Yuan, and Ying Shan.
\newblock Dreamdiffusion: Generating high-quality images from brain eeg signals, 2023.

\bibitem{Ballini2014A}
M.~Ballini, J.~Muller, P.~Livi, Yihui Chen, U.~Frey, A.~Stettler, A.~Shadmani, Vijay Viswam, Ian~Lloyd Jones, D.~Jackel, M.~Radivojevic, M.~Lewandowska, W.~Gong, M.~Fiscella, D.~Bakkum, F.~Heer, and A.~Hierlemann.
\newblock A 1024-channel cmos microelectrode array with 26,400 electrodes for recording and stimulation of electrogenic cells in vitro.
\newblock {\em IEEE Journal of Solid-State Circuits}, 49:2705--2719, 2014.

\bibitem{blattmann2023stable}
Andreas Blattmann, Tim Dockhorn, Sumith Kulal, Daniel Mendelevitch, Maciej Kilian, Dominik Lorenz, Yam Levi, Zion English, Vikram Voleti, Adam Letts, et~al.
\newblock Stable video diffusion: Scaling latent video diffusion models to large datasets.
\newblock {\em arXiv preprint arXiv:2311.15127}, 2023.

\bibitem{Bollimunta2020Head-mounted}
Anil Bollimunta, Samantha~R. Santacruz, Ryan~W. Eaton, P.~S. Xu, J.~Morrison, K.~Moxon, J.~Carmena, and J.~Nassi.
\newblock Head-mounted microendoscopic calcium imaging in dorsal premotor cortex of behaving rhesus macaque.
\newblock {\em Cell reports}, 35:109239 -- 109239, 2020.

\bibitem{Cadena2017DeepCM}
Santiago~A. Cadena, George~H. Denfield, Edgar~Y. Walker, Leon~A. Gatys, Andreas~Savas Tolias, Matthias Bethge, and Alexander~S. Ecker.
\newblock Deep convolutional models improve predictions of macaque v1 responses to natural images.
\newblock {\em PLoS Computational Biology}, 15, 2017.

\bibitem{chen2023iquery}
Jiaben Chen, Renrui Zhang, Dongze Lian, Jiaqi Yang, Ziyao Zeng, and Jianbo Shi.
\newblock iquery: Instruments as queries for audio-visual sound separation.
\newblock In {\em Proceedings of the IEEE/CVF Conference on Computer Vision and Pattern Recognition}, pages 14675--14686, 2023.

\bibitem{chen2020simple}
Ting Chen, Simon Kornblith, Mohammad Norouzi, and Geoffrey Hinton.
\newblock A simple framework for contrastive learning of visual representations, 2020.

\bibitem{Chen2023CinematicMH}
Zijiao Chen, Jiaxin Qing, and Juan~Helen Zhou.
\newblock Cinematic mindscapes: High-quality video reconstruction from brain activity.
\newblock {\em ArXiv}, abs/2305.11675, 2023.

\bibitem{Jong2001CanonicalPL}
Sijmen de~Jong, Barry~M. Wise, and N.~L. Ricker.
\newblock Canonical partial least squares and continuum power regression.
\newblock {\em Journal of Chemometrics}, 15, 2001.

\bibitem{dosovitskiy2021an}
Alexey Dosovitskiy, Lucas Beyer, Alexander Kolesnikov, Dirk Weissenborn, Xiaohua Zhai, Thomas Unterthiner, Mostafa Dehghani, Matthias Minderer, Georg Heigold, Sylvain Gelly, Jakob Uszkoreit, and Neil Houlsby.
\newblock An image is worth 16x16 words: Transformers for image recognition at scale.
\newblock In {\em International Conference on Learning Representations}, 2021.

\bibitem{dou2024tactile}
Yiming Dou, Fengyu Yang, Yi~Liu, Antonio Loquercio, and Andrew Owens.
\newblock Tactile-augmented radiance fields.
\newblock In {\em Proceedings of the IEEE/CVF Conference on Computer Vision and Pattern Recognition}, 2024.

\bibitem{Du2009High-resolution}
Jiangang Du, I.~Riedel-Kruse, J.~Nawroth, M.~Roukes, G.~Laurent, and S.~Masmanidis.
\newblock High-resolution three-dimensional extracellular recording of neuronal activity with microfabricated electrode arrays.
\newblock {\em Journal of neurophysiology}, 101 3:1671--8, 2009.

\bibitem{feng2023self}
Chao Feng, Ziyang Chen, and Andrew Owens.
\newblock Self-supervised video forensics by audio-visual anomaly detection.
\newblock In {\em Proceedings of the IEEE/CVF Conference on Computer Vision and Pattern Recognition}, pages 10491--10503, 2023.

\bibitem{festa2021neuronal}
Dylan Festa, Amir Aschner, Aida Davila, Adam Kohn, and Ruben Coen-Cagli.
\newblock Neuronal variability reflects probabilistic inference tuned to natural image statistics.
\newblock {\em Nature communications}, 12(1):3635, 2021.

\bibitem{gao2023llama}
Peng Gao, Jiaming Han, Renrui Zhang, Ziyi Lin, Shijie Geng, Aojun Zhou, Wei Zhang, Pan Lu, Conghui He, Xiangyu Yue, et~al.
\newblock Llama-adapter v2: Parameter-efficient visual instruction model.
\newblock {\em arXiv preprint arXiv:2304.15010}, 2023.

\bibitem{Girdhar2023ImageBindOE}
Rohit Girdhar, Alaaeldin El-Nouby, Zhuang Liu, Mannat Singh, Kalyan~Vasudev Alwala, Armand Joulin, and Ishan Misra.
\newblock Imagebind: One embedding space to bind them all.
\newblock {\em 2023 IEEE/CVF Conference on Computer Vision and Pattern Recognition (CVPR)}, pages 15180--15190, 2023.

\bibitem{han2023onellm}
Jiaming Han, Kaixiong Gong, Yiyuan Zhang, Jiaqi Wang, Kaipeng Zhang, Dahua Lin, Yu~Qiao, Peng Gao, and Xiangyu Yue.
\newblock Onellm: One framework to align all modalities with language, 2023.

\bibitem{han2023imagebind}
Jiaming Han, Renrui Zhang, Wenqi Shao, Peng Gao, Peng Xu, Han Xiao, Kaipeng Zhang, Chris Liu, Song Wen, Ziyu Guo, et~al.
\newblock Imagebind-llm: Multi-modality instruction tuning.
\newblock {\em arXiv preprint arXiv:2309.03905}, 2023.

\bibitem{heusel2017gans}
Martin Heusel, Hubert Ramsauer, Thomas Unterthiner, Bernhard Nessler, and Sepp Hochreiter.
\newblock Gans trained by a two time-scale update rule converge to a local nash equilibrium.
\newblock {\em Advances in neural information processing systems}, 30, 2017.

\bibitem{ho2020denoising}
Jonathan Ho, Ajay Jain, and Pieter Abbeel.
\newblock Denoising diffusion probabilistic models.
\newblock {\em Advances in neural information processing systems}, 33:6840--6851, 2020.

\bibitem{ho2022video}
Jonathan Ho, Tim Salimans, Alexey Gritsenko, William Chan, Mohammad Norouzi, and David~J Fleet.
\newblock Video diffusion models.
\newblock {\em Advances in Neural Information Processing Systems}, 35:8633--8646, 2022.

\bibitem{Horikawa2015GenericDO}
Tomoyasu Horikawa and Yukiyasu Kamitani.
\newblock Generic decoding of seen and imagined objects using hierarchical visual features.
\newblock {\em Nature Communications}, 8, 2015.

\bibitem{Hotelling1936RelationsBT}
Harold Hotelling.
\newblock Relations between two sets of variates.
\newblock {\em Biometrika}, 28:321--377, 1936.

\bibitem{Huang2020PerceptiontoImageRN}
Wei Huang, Hongmei Yan, Chong Wang, Jiyi Li, Zhentao Zuo, Jiang Zhang, Zhan Shen, and Huafu Chen.
\newblock Perception-to-image: Reconstructing natural images from the brain activity of visual perception.
\newblock {\em Annals of Biomedical Engineering}, 48:2323 -- 2332, 2020.

\bibitem{Kavasidis2017Brain2ImageCB}
Isaak Kavasidis, Simone Palazzo, Concetto Spampinato, Daniela Giordano, and Mubarak Shah.
\newblock Brain2image: Converting brain signals into images.
\newblock {\em Proceedings of the 25th ACM international conference on Multimedia}, 2017.

\bibitem{kay2008identifying}
Kendrick~N Kay, Thomas Naselaris, Ryan~J Prenger, and Jack~L Gallant.
\newblock Identifying natural images from human brain activity.
\newblock {\em Nature}, 452(7185):352--355, 2008.

\bibitem{Kell2018ATN}
Alexander J.~E. Kell, Daniel Yamins, Erica~N. Shook, Sam~V. Norman-Haignere, and Josh~H. McDermott.
\newblock A task-optimized neural network replicates human auditory behavior, predicts brain responses, and reveals a cortical processing hierarchy.
\newblock {\em Neuron}, 98:630--644.e16, 2018.

\bibitem{khosla2021cortical}
Meenakshi Khosla, Gia~H Ngo, Keith Jamison, Amy Kuceyeski, and Mert~R Sabuncu.
\newblock Cortical response to naturalistic stimuli is largely predictable with deep neural networks.
\newblock {\em Science Advances}, 7(22):eabe7547, 2021.

\bibitem{kingma2015adam}
Diederik Kingma and Jimmy Ba.
\newblock Adam: A method for stochastic optimization.
\newblock In {\em International Conference on Learning Representation}, 2015.

\bibitem{Wonjun_EEG_representation}
Wonjun Ko, Eunjin Jeon, Seungwoo Jeong, and Heung-Il Suk.
\newblock Multi-scale neural network for eeg representation learning in bci.
\newblock {\em IEEE Computational Intelligence Magazine}, 16(2):31--45, 2021.

\bibitem{kohn2005stimulus}
Adam Kohn and Matthew~A Smith.
\newblock Stimulus dependence of neuronal correlation in primary visual cortex of the macaque.
\newblock {\em Journal of Neuroscience}, 25(14):3661--3673, 2005.

\bibitem{lao2022depth}
Dong Lao, Fengyu Yang, Daniel Wang, Hyoungseob Park, Samuel Lu, Alex Wong, and Stefano Soatto.
\newblock On the viability of monocular depth pre-training for semantic segmentation.
\newblock {\em arXiv preprint arXiv:2203.13987}, 2022.

\bibitem{lee2022sound}
Seung~Hyun Lee, Wonseok Roh, Wonmin Byeon, Sang~Ho Yoon, Chanyoung Kim, Jinkyu Kim, and Sangpil Kim.
\newblock Sound-guided semantic image manipulation.
\newblock In {\em Proceedings of the IEEE/CVF conference on computer vision and pattern recognition}, pages 3377--3386, 2022.

\bibitem{Lei2023ViTLens2GT}
Weixian Lei, Yixiao Ge, Kun Yi, Jianfeng Zhang, Difei Gao, Dylan Sun, Yuying Ge, Ying Shan, and Mike~Zheng Shou.
\newblock Vit-lens-2: Gateway to omni-modal intelligence.
\newblock {\em ArXiv}, abs/2311.16081, 2023.

\bibitem{li-23-deception-detection}
Panfeng Li, Mohamed Abouelenien, and Rada Mihalcea.
\newblock Deception detection from linguistic and physiological data streams using bimodal convolutional neural networks.
\newblock {\em arXiv preprint arXiv:2311.10944}, 2023.

\bibitem{li-19-segmentation}
Panfeng Li, Youzuo Lin, and Emily Schultz-Fellenz.
\newblock Contextual hourglass network for semantic segmentation of high resolution aerial imagery.
\newblock {\em arXiv preprint arXiv:1810.12813}, 2019.

\bibitem{li-24-vqa}
Panfeng Li, Qikai Yang, Xieming Geng, Wenjing Zhou, Zhicheng Ding, and Yi~Nian.
\newblock Exploring diverse methods in visual question answering.
\newblock {\em arXiv preprint arXiv:2404.13565}, 2024.

\bibitem{li2023dissecting}
Yuanning Li, Gopala~K Anumanchipalli, Abdelrahman Mohamed, Peili Chen, Laurel~H Carney, Junfeng Lu, Jinsong Wu, and Edward~F Chang.
\newblock Dissecting neural computations in the human auditory pathway using deep neural networks for speech.
\newblock {\em Nature Neuroscience}, 26(12):2213--2225, 2023.

\bibitem{lin2004rouge}
Chin-Yew Lin.
\newblock Rouge: A package for automatic evaluation of summaries.
\newblock In {\em Text summarization branches out}, pages 74--81, 2004.

\bibitem{lin2024optimalsamplinglearningsdf}
Guying Lin, Lei Yang, Yuan Liu, Congyi Zhang, Junhui Hou, Xiaogang Jin, Taku Komura, John Keyser, and Wenping Wang.
\newblock On optimal sampling for learning sdf using mlps equipped with positional encoding, 2024.

\bibitem{lin2023patch}
Guying Lin, Lei Yang, Congyi Zhang, Hao Pan, Yuhan Ping, Guodong Wei, Taku Komura, John Keyser, and Wenping Wang.
\newblock Patch-grid: an efficient and feature-preserving neural implicit surface representation.
\newblock {\em arXiv preprint arXiv:2308.13934}, 2023.

\bibitem{lin2022mind}
Sikun Lin, Thomas Sprague, and Ambuj~K Singh.
\newblock Mind reader: Reconstructing complex images from brain activities, 2022.

\bibitem{lin2015microsoft}
Tsung-Yi Lin, Michael Maire, Serge Belongie, Lubomir Bourdev, Ross Girshick, James Hays, Pietro Perona, Deva Ramanan, C.~Lawrence Zitnick, and Piotr Dollár.
\newblock Microsoft coco: Common objects in context, 2015.

\bibitem{liu2023brainclip}
Yulong Liu, Yongqiang Ma, Wei Zhou, Guibo Zhu, and Nanning Zheng.
\newblock Brainclip: Bridging brain and visual-linguistic representation via clip for generic natural visual stimulus decoding, 2023.

\bibitem{loshchilov2017decoupled}
Ilya Loshchilov and Frank Hutter.
\newblock Decoupled weight decay regularization.
\newblock {\em arXiv preprint arXiv:1711.05101}, 2017.

\bibitem{Lu2023MindDiffuserCI}
Yizhuo Lu, Changde Du, Qiongyi Zhou, Dianpeng Wang, and Huiguang He.
\newblock Minddiffuser: Controlled image reconstruction from human brain activity with semantic and structural diffusion.
\newblock {\em Proceedings of the 31st ACM International Conference on Multimedia}, 2023.

\bibitem{mai2024brainconditional}
Weijian Mai, Jian Zhang, Pengfei Fang, and Zhijun Zhang.
\newblock Brain-conditional multimodal synthesis: A survey and taxonomy, 2024.

\bibitem{Martin2013Functional}
K.~Martin and Sylvia Schröder.
\newblock Functional heterogeneity in neighboring neurons of cat primary visual cortex in response to both artificial and natural stimuli.
\newblock {\em The Journal of Neuroscience}, 33:7325 -- 7344, 2013.

\bibitem{mcinnes2018umap}
Leland McInnes, John Healy, and James Melville.
\newblock Umap: Uniform manifold approximation and projection for dimension reduction.
\newblock {\em arXiv preprint arXiv:1802.03426}, 2018.

\bibitem{millet2022toward}
Juliette Millet, Charlotte Caucheteux, Yves Boubenec, Alexandre Gramfort, Ewan Dunbar, Christophe Pallier, Jean-Remi King, et~al.
\newblock Toward a realistic model of speech processing in the brain with self-supervised learning.
\newblock {\em Advances in Neural Information Processing Systems}, 35:33428--33443, 2022.

\bibitem{Mishra2022NeuroGANIR}
Rahul Mishra, Krishan Sharma, Ranjeet~Ranjan Jha, and Arnav~V. Bhavsar.
\newblock Neurogan: image reconstruction from eeg signals via an attention-based gan.
\newblock {\em Neural Computing and Applications}, 35:9181--9192, 2022.

\bibitem{nishimoto2011reconstructing}
Shinji Nishimoto, An~T Vu, Thomas Naselaris, Yuval Benjamini, Bin Yu, and Jack~L Gallant.
\newblock Reconstructing visual experiences from brain activity evoked by natural movies.
\newblock {\em Current biology}, 21(19):1641--1646, 2011.

\bibitem{oord2018representation}
Aaron van~den Oord, Yazhe Li, and Oriol Vinyals.
\newblock Representation learning with contrastive predictive coding.
\newblock {\em arXiv preprint arXiv:1807.03748}, 2018.

\bibitem{ozcelik2023natural}
Furkan Ozcelik and Rufin VanRullen.
\newblock Natural scene reconstruction from fmri signals using generative latent diffusion.
\newblock {\em Scientific Reports}, 13(1):15666, 2023.

\bibitem{palazzo2020decoding}
Simone Palazzo, Concetto Spampinato, Isaak Kavasidis, Daniela Giordano, Joseph Schmidt, and Mubarak Shah.
\newblock Decoding brain representations by multimodal learning of neural activity and visual features, 2020.

\bibitem{Panzeri2008On}
S.~Panzeri, C.~Magri, and N.~Logothetis.
\newblock On the use of information theory for the analysis of the relationship between neural and imaging signals.
\newblock {\em Magnetic resonance imaging}, 26 7:1015--25, 2008.

\bibitem{poole2022dreamfusion}
Ben Poole, Ajay Jain, Jonathan~T Barron, and Ben Mildenhall.
\newblock Dreamfusion: Text-to-3d using 2d diffusion.
\newblock {\em arXiv preprint arXiv:2209.14988}, 2022.

\bibitem{qiu2021vt}
Longtian Qiu, Renrui Zhang, Ziyu Guo, Ziyao Zeng, Zilu Guo, Yafeng Li, and Guangnan Zhang.
\newblock Vt-clip: Enhancing vision-language models with visual-guided texts.
\newblock {\em arXiv preprint arXiv:2112.02399}, 2021.

\bibitem{radford2021learning}
Alec Radford, Jong~Wook Kim, Chris Hallacy, Aditya Ramesh, Gabriel Goh, Sandhini Agarwal, Girish Sastry, Amanda Askell, Pamela Mishkin, Jack Clark, Gretchen Krueger, and Ilya Sutskever.
\newblock Learning transferable visual models from natural language supervision, 2021.

\bibitem{ramesh2022hierarchical}
Aditya Ramesh, Prafulla Dhariwal, Alex Nichol, Casey Chu, and Mark Chen.
\newblock Hierarchical text-conditional image generation with clip latents.
\newblock {\em arXiv preprint arXiv:2204.06125}, 1(2):3, 2022.

\bibitem{rombach2022high}
Robin Rombach, Andreas Blattmann, Dominik Lorenz, Patrick Esser, and Bj{\"o}rn Ommer.
\newblock High-resolution image synthesis with latent diffusion models.
\newblock In {\em Proceedings of the IEEE/CVF conference on computer vision and pattern recognition}, pages 10684--10695, 2022.

\bibitem{Sadakane2015Long-Term}
O.~Sadakane, Yoshito Masamizu, A.~Watakabe, Shin-Ichiro Terada, Masanari Ohtsuka, Masafumi Takaji, H.~Mizukami, K.~Ozawa, H.~Kawasaki, M.~Matsuzaki, and T.~Yamamori.
\newblock Long-term two-photon calcium imaging of neuronal populations with subcellular resolution in adult non-human primates.
\newblock {\em Cell reports}, 13 9:1989--99, 2015.

\bibitem{salimans2016improved}
Tim Salimans, Ian Goodfellow, Wojciech Zaremba, Vicki Cheung, Alec Radford, and Xi~Chen.
\newblock Improved techniques for training gans.
\newblock {\em Advances in neural information processing systems}, 29, 2016.

\bibitem{sani2021modeling}
Omid~G Sani, Hamidreza Abbaspourazad, Yan~T Wong, Bijan Pesaran, and Maryam~M Shanechi.
\newblock Modeling behaviorally relevant neural dynamics enabled by preferential subspace identification.
\newblock {\em Nature Neuroscience}, 24(1):140--149, 2021.

\bibitem{Schneider_2023}
Steffen Schneider, Jin~Hwa Lee, and Mackenzie~Weygandt Mathis.
\newblock Learnable latent embeddings for joint behavioural and neural analysis.
\newblock {\em Nature}, 617(7960):360–368, May 2023.

\bibitem{Scotti2023ReconstructingTM}
Paul~S. Scotti, Atmadeep Banerjee, Jimmie Goode, Stepan Shabalin, Alex Nguyen, Ethan Cohen, Aidan Dempster, Nathalie Verlinde, Elad Yundler, David Weisberg, Kenneth~A. Norman, and T.~Abraham.
\newblock Reconstructing the mind's eye: fmri-to-image with contrastive learning and diffusion priors.
\newblock {\em ArXiv}, abs/2305.18274, 2023.

\bibitem{seeliger2018generative}
Katja Seeliger, Umut G{\"u}{\c{c}}l{\"u}, Luca Ambrogioni, Yagmur G{\"u}{\c{c}}l{\"u}t{\"u}rk, and Marcel~AJ Van~Gerven.
\newblock Generative adversarial networks for reconstructing natural images from brain activity.
\newblock {\em NeuroImage}, 181:775--785, 2018.

\bibitem{song2020denoising}
Jiaming Song, Chenlin Meng, and Stefano Ermon.
\newblock Denoising diffusion implicit models.
\newblock {\em arXiv preprint arXiv:2010.02502}, 2020.

\bibitem{Tajbakhsh2016Convolutional}
Nima Tajbakhsh, Jae~Y. Shin, S.~Gurudu, R.~T. Hurst, Christopher~B. Kendall, M.~Gotway, and Jianming Liang.
\newblock Convolutional neural networks for medical image analysis: Full training or fine tuning?
\newblock {\em IEEE Transactions on Medical Imaging}, 35:1299--1312, 2016.

\bibitem{Takagi2023HighresolutionIR}
Yu~Takagi and Shinji Nishimoto.
\newblock High-resolution image reconstruction with latent diffusion models from human brain activity.
\newblock {\em bioRxiv}, 2023.

\bibitem{takagi2023high}
Yu~Takagi and Shinji Nishimoto.
\newblock High-resolution image reconstruction with latent diffusion models from human brain activity.
\newblock In {\em Proceedings of the IEEE/CVF Conference on Computer Vision and Pattern Recognition}, pages 14453--14463, 2023.

\bibitem{tang2023semantic}
Jerry Tang, Amanda LeBel, Shailee Jain, and Alexander~G Huth.
\newblock Semantic reconstruction of continuous language from non-invasive brain recordings.
\newblock {\em Nature Neuroscience}, 26(5):858--866, 2023.

\bibitem{touvron2023llama}
Hugo Touvron, Thibaut Lavril, Gautier Izacard, Xavier Martinet, Marie-Anne Lachaux, Timoth{\'e}e Lacroix, Baptiste Rozi{\`e}re, Naman Goyal, Eric Hambro, Faisal Azhar, et~al.
\newblock Llama: Open and efficient foundation language models.
\newblock {\em arXiv preprint arXiv:2302.13971}, 2023.

\bibitem{Touvron2023Llama2O}
Hugo Touvron, Louis Martin, Kevin~R. Stone, Peter Albert, Amjad Almahairi, Yasmine Babaei, Nikolay Bashlykov, Soumya Batra, Prajjwal Bhargava, Shruti Bhosale, Daniel~M. Bikel, Lukas Blecher, Cristian~Cant{\'o}n Ferrer, Moya Chen, Guillem Cucurull, David Esiobu, Jude Fernandes, Jeremy Fu, Wenyin Fu, Brian Fuller, Cynthia Gao, Vedanuj Goswami, Naman Goyal, Anthony~S. Hartshorn, Saghar Hosseini, Rui Hou, Hakan Inan, Marcin Kardas, Viktor Kerkez, Madian Khabsa, Isabel~M. Kloumann, A.~V. Korenev, Punit~Singh Koura, Marie-Anne Lachaux, Thibaut Lavril, Jenya Lee, Diana Liskovich, Yinghai Lu, Yuning Mao, Xavier Martinet, Todor Mihaylov, Pushkar Mishra, Igor Molybog, Yixin Nie, Andrew Poulton, Jeremy Reizenstein, Rashi Rungta, Kalyan Saladi, Alan Schelten, Ruan Silva, Eric~Michael Smith, R.~Subramanian, Xia Tan, Binh Tang, Ross Taylor, Adina Williams, Jian~Xiang Kuan, Puxin Xu, Zhengxu Yan, Iliyan Zarov, Yuchen Zhang, Angela Fan, Melanie Kambadur, Sharan Narang, Aurelien Rodriguez, Robert Stojnic, Sergey Edunov, and
  Thomas Scialom.
\newblock Llama 2: Open foundation and fine-tuned chat models.
\newblock {\em ArXiv}, abs/2307.09288, 2023.

\bibitem{uludaug2014general}
K{\^a}mil Uluda{\u{g}} and Alard Roebroeck.
\newblock General overview on the merits of multimodal neuroimaging data fusion.
\newblock {\em Neuroimage}, 102:3--10, 2014.

\bibitem{urai2022large}
Anne~E Urai, Brent Doiron, Andrew~M Leifer, and Anne~K Churchland.
\newblock Large-scale neural recordings call for new insights to link brain and behavior.
\newblock {\em Nature neuroscience}, 25(1):11--19, 2022.

\bibitem{van2008visualizing}
Laurens Van~der Maaten and Geoffrey Hinton.
\newblock Visualizing data using t-sne.
\newblock {\em Journal of machine learning research}, 9(11), 2008.

\bibitem{vedantam2015cider}
Ramakrishna Vedantam, C~Lawrence~Zitnick, and Devi Parikh.
\newblock Cider: Consensus-based image description evaluation.
\newblock In {\em Proceedings of the IEEE conference on computer vision and pattern recognition}, pages 4566--4575, 2015.

\bibitem{wang2023large}
Tianye Wang, Haoxuan Yao, Tai~Sing Lee, Jiayi Hong, Yang Li, Hongfei Jiang, Ian~Max Andolina, and Shiming Tang.
\newblock A large calcium-imaging dataset reveals a systematic v4 organization for natural scenes, 2023.

\bibitem{Wen2016NeuralEA}
Haiguang Wen, Junxing Shi, Yizhen Zhang, Kun-Han Lu, Jiayu Cao, and Zhongming Liu.
\newblock Neural encoding and decoding with deep learning for dynamic natural vision.
\newblock {\em Cerebral Cortex}, 28:4136–4160, 2016.

\bibitem{Whittingstall2013Physiological}
Kevin Whittingstall and Nikos~K Logothetis.
\newblock Physiological foundations of neural signals.
\newblock {\em Principles of Neural Coding}, 15:146, 2013.

\bibitem{Wu_2023_boosting}
Shaokai Wu and Fengyu Yang.
\newblock Boosting detection in crowd analysis via underutilized output features.
\newblock In {\em Proceedings of the IEEE/CVF Conference on Computer Vision and Pattern Recognition (CVPR)}, pages 15609--15618, June 2023.

\bibitem{wu2018unsupervised}
Zhirong Wu, Yuanjun Xiong, Stella~X Yu, and Dahua Lin.
\newblock Unsupervised feature learning via non-parametric instance discrimination.
\newblock In {\em Proceedings of the IEEE conference on computer vision and pattern recognition}, pages 3733--3742, 2018.

\bibitem{xu2024vision}
Zhiyang Xu, Chao Feng, Rulin Shao, Trevor Ashby, Ying Shen, Di~Jin, Yu~Cheng, Qifan Wang, and Lifu Huang.
\newblock Vision-flan: Scaling human-labeled tasks in visual instruction tuning.
\newblock {\em arXiv preprint arXiv:2402.11690}, 2024.

\bibitem{Xue2022ULIPLA}
Le~Xue, Mingfei Gao, Chen Xing, Roberto Mart'in-Mart'in, Jiajun Wu, Caiming Xiong, Ran Xu, Juan~Carlos Niebles, and Silvio Savarese.
\newblock Ulip: Learning a unified representation of language, images, and point clouds for 3d understanding.
\newblock {\em 2023 IEEE/CVF Conference on Computer Vision and Pattern Recognition (CVPR)}, pages 1179--1189, 2022.

\bibitem{yamins2013hierarchical}
Daniel~L Yamins, Ha~Hong, Charles Cadieu, and James~J DiCarlo.
\newblock Hierarchical modular optimization of convolutional networks achieves representations similar to macaque it and human ventral stream.
\newblock {\em Advances in neural information processing systems}, 26, 2013.

\bibitem{yamins2016using}
Daniel~LK Yamins and James~J DiCarlo.
\newblock Using goal-driven deep learning models to understand sensory cortex.
\newblock {\em Nature neuroscience}, 19(3):356--365, 2016.

\bibitem{yang2024binding}
Fengyu Yang, Chao Feng, Ziyang Chen, Hyoungseob Park, Daniel Wang, Yiming Dou, Ziyao Zeng, Xien Chen, Rit Gangopadhyay, Andrew Owens, et~al.
\newblock Binding touch to everything: Learning unified multimodal tactile representations.
\newblock In {\em Proceedings of the IEEE/CVF Conference on Computer Vision and Pattern Recognition}, pages 26340--26353, 2024.

\bibitem{yang2022sparse}
Fengyu Yang and Chenyang Ma.
\newblock Sparse and complete latent organization for geospatial semantic segmentation.
\newblock In {\em Proceedings of the IEEE/CVF Conference on Computer Vision and Pattern Recognition}, pages 1809--1818, 2022.

\bibitem{yang2022touch}
Fengyu Yang, Chenyang Ma, Jiacheng Zhang, Jing Zhu, Wenzhen Yuan, and Andrew Owens.
\newblock Touch and go: Learning from human-collected vision and touch.
\newblock {\em Neural Information Processing Systems (NeurIPS) - Datasets and Benchmarks Track}, 2022.

\bibitem{yang2023generating}
Fengyu Yang, Jiacheng Zhang, and Andrew Owens.
\newblock Generating visual scenes from touch.
\newblock {\em International Conference on Computer Vision (ICCV)}, 2023.

\bibitem{Yoshida2020NaturalIA}
Takashi Yoshida and Kenichi Ohki.
\newblock Natural images are reliably represented by sparse and variable populations of neurons in visual cortex.
\newblock {\em Nature Communications}, 11, 2020.

\bibitem{zeng2024wordepth}
Ziyao Zeng, Daniel Wang, Fengyu Yang, Hyoungseob Park, Stefano Soatto, Dong Lao, and Alex Wong.
\newblock Wordepth: Variational language prior for monocular depth estimation.
\newblock In {\em Proceedings of the IEEE/CVF Conference on Computer Vision and Pattern Recognition}, pages 9708--9719, 2024.

\bibitem{zhang2023llama}
Renrui Zhang, Jiaming Han, Aojun Zhou, Xiangfei Hu, Shilin Yan, Pan Lu, Hongsheng Li, Peng Gao, and Yu~Qiao.
\newblock Llama-adapter: Efficient fine-tuning of language models with zero-init attention.
\newblock {\em arXiv preprint arXiv:2303.16199}, 2023.

\bibitem{zhang2021dspoint}
Renrui Zhang, Ziyao Zeng, Ziyu Guo, Xinben Gao, Kexue Fu, and Jianbo Shi.
\newblock Dspoint: Dual-scale point cloud recognition with high-frequency fusion.
\newblock {\em arXiv preprint arXiv:2111.10332}, 2021.

\bibitem{zhang2022can}
Renrui Zhang, Ziyao Zeng, Ziyu Guo, and Yafeng Li.
\newblock Can language understand depth?
\newblock In {\em Proceedings of the 30th ACM International Conference on Multimedia}, pages 6868--6874, 2022.

\bibitem{zhang2016colorful}
Richard Zhang, Phillip Isola, and Alexei~A Efros.
\newblock Colorful image colorization.
\newblock In {\em ECCV}, 2016.

\bibitem{zhang2017real}
Richard Zhang, Jun-Yan Zhu, Phillip Isola, Xinyang Geng, Angela~S Lin, Tianhe Yu, and Alexei~A Efros.
\newblock Real-time user-guided image colorization with learned deep priors.
\newblock {\em ACM Transactions on Graphics (TOG)}, 9(4), 2017.

\bibitem{Zhao2022RBCRT}
Hanbin Zhao, Fengyu Yang, Xinghe Fu, and Xi~Li.
\newblock Rbc: Rectifying the biased context in continual semantic segmentation.
\newblock {\em ECCV}, 2022.

\bibitem{Zhen2019DeepSC}
Liangli Zhen, Peng Hu, Xu~Wang, and Dezhong Peng.
\newblock Deep supervised cross-modal retrieval.
\newblock {\em 2019 IEEE/CVF Conference on Computer Vision and Pattern Recognition (CVPR)}, pages 10386--10395, 2019.

\bibitem{zheng2021unraveling}
Yajing Zheng, Shanshan Jia, Zhaofei Yu, Jian~K Liu, and Tiejun Huang.
\newblock Unraveling neural coding of dynamic natural visual scenes via convolutional recurrent neural networks.
\newblock {\em Patterns}, 2(10), 2021.

\bibitem{zhou2020learning}
Ding Zhou and Xue-Xin Wei.
\newblock Learning identifiable and interpretable latent models of high-dimensional neural activity using pi-vae.
\newblock {\em Advances in Neural Information Processing Systems}, 33:7234--7247, 2020.

\bibitem{Zhou2024CLIPMUSEDCM}
Qiongyi Zhou, Changde Du, Shengpei Wang, and Huiguang He.
\newblock Clip-mused: Clip-guided multi-subject visual neural information semantic decoding.
\newblock {\em ArXiv}, abs/2402.08994, 2024.

\bibitem{zhu2023pointclip}
Xiangyang Zhu, Renrui Zhang, Bowei He, Ziyu Guo, Ziyao Zeng, Zipeng Qin, Shanghang Zhang, and Peng Gao.
\newblock Pointclip v2: Prompting clip and gpt for powerful 3d open-world learning.
\newblock In {\em Proceedings of the IEEE/CVF International Conference on Computer Vision}, pages 2639--2650, 2023.

\end{thebibliography}
